\pdfoutput=1

\newcommand{\colswitch}[2]{#2}

\documentclass[prd,superscriptaddress,nofootinbib, floatfix, \colswitch{}{twocolumn}]{revtex4}
\usepackage{bm,amsmath,amssymb,epsfig,natbib}
\usepackage{color}
\usepackage{ifthen}

\begin{document}

\renewcommand{\vec}[1]{\mathbf{#1}}
\newcommand{\panel}[1]{\emph{#1}:}
\newcommand{\ad}{\text{Ad}}
\newcommand{\iso}{\text{Iso}}

\newcommand{\lexact}{{l_\text{exact}}}
\newcommand{\llow}{{l_\text{low}}}
\newcommand{\Ctot}{C^{\rm{tot}}}

\newcommand{\Li}{\text{Li}}

\newcommand {\mDh} { \hat{\bm{D} } }
\newcommand {\mUh} { \hat{\bm{U} } }

\newcommand{\nside}{N_{\text{side}}}
\newcommand{\arcmin}{\text{arcmin}}
\newcommand{\begm}{\begin{pmatrix}}
\newcommand{\enm}{\end{pmatrix}}
\newcommand{\threej}[6]{{\begm #1 & #2 & #3 \\ #4 & #5 & #6 \enm}}
\newcommand{\threejz}[3]{{\begm #1 & #2 & #3 \\ 0 & 0 & 0 \enm}}

\newcommand{\Coff}{{\hat{C}_l^\off}}
\newcommand{\lmax}{{l_\text{max}}}
\newcommand{\lmin}{l_{\text{min}}}
\newcommand{\fsky}{f_{\text{sky}}}
\newcommand{\off}{{\text{off}}}
\newcommand{\chieff}{\chi^2_{\text{eff}}}
\newcommand{\nHI}{{n_{HI}}}
\newcommand{\clh}{\mathcal{H}}
\newcommand{\ud}{{\rm d}}

\def\eprinttmp@#1arXiv:#2 [#3]#4@{
\ifthenelse{\equal{#3}{x}}{\href{http://arxiv.org/abs/#1}{#1}}{\href{http://arxiv.org/abs/#2}{arXiv:#2} [#3]}}

\renewcommand{\eprint}[1]{\eprinttmp@#1arXiv: [x]@}
\newcommand{\adsurl}[1]{\href{#1}{ADS}}
\renewcommand{\bibinfo}[2]{\ifthenelse{\equal{#1}{isbn}}{
\href{http://cosmologist.info/ISBN/#2}{#2}}{#2}}

\newcommand{\cle}{\mathcal{E}}
\newcommand{\clp}{\mathcal{P}}
\newcommand{\bTheta}{\bar{\Theta}}

\newcommand\ba{\begin{eqnarray}}
\newcommand\ea{\end{eqnarray}}
\newcommand\be{\begin{equation}}
\newcommand\ee{\end{equation}}
\newcommand\lagrange{{\cal L}}
\newcommand\cll{{\cal L}}
\newcommand\clx{{\cal X}}
\newcommand\clz{{\cal Z}}
\newcommand\clv{{\cal V}}
\newcommand\clo{{\cal O}}
\newcommand\cla{{\cal A}}
\newcommand{\uD}{{\mathrm{D}}}
\newcommand{\calE}{{\cal E}}
\newcommand{\calB}{{\cal B}}
\newcommand{\curl}{\,\mbox{curl}\,}
\newcommand\del{\nabla}
\newcommand\Tr{{\rm Tr}}
\newcommand\half{{\frac{1}{2}}}
\renewcommand\H{{\cal H}}
\newcommand\K{{\rm K}}
\newcommand\mK{{\rm mK}}
\newcommand{\clk}{{\cal K}}
\newcommand{\bq}{\bar{q}}
\newcommand{\bv}{\bar{v}}
\renewcommand\P{{\cal P}}
\newcommand{\numfrac}[2]{{\textstyle \frac{#1}{#2}}}
\newcommand{\la}{\langle}
\newcommand{\ra}{\rangle}
\newcommand{\rar}{\rightarrow}
\newcommand{\Rar}{\Rightarrow}
\newcommand\gsim{ \lower .75ex \hbox{$\sim$} \llap{\raise .27ex \hbox{$>$}} }
\newcommand\lsim{ \lower .75ex \hbox{$\sim$} \llap{\raise .27ex \hbox{$<$}} }
\newcommand\bigdot[1] {\stackrel{\mbox{{\huge .}}}{#1}}
\newcommand\bigddot[1] {\stackrel{\mbox{{\huge ..}}}{#1}}
\newcommand{\Mpc}{\text{Mpc}}
\newcommand{\Al}{{A_l}}
\newcommand{\Bl}{{B_l}}
\newcommand{\eAl}{e^\Al}
\newcommand{\ix}{{(i)}}
\newcommand{\ixp}{{(i+1)}}
\renewcommand{\k}{\beta}
\newcommand{\HD}{\mathrm{D}}
\newcommand{\mCh}{\hat{\bm{C}}}
\newcommand{\mCf}{{{\bm{C}}_{f}}}
\newcommand{\mCXf}{{{\bm{C}}_{Xf}}}
\newcommand{\mMXf}{{{\bm{M}}_{f}}}
\newcommand{\Cfl}{{C_f}_l}
\newcommand{\muK}{\mu \rm{K}}

\newcommand{\Xh}{\hat{X}}

\newcommand{\var}{\text{var}}
\newcommand{\cov}{\text{cov}}
\newcommand{\bias}{\text{bias}}

\newcommand{\vecp}{\text{vecp}}
\newcommand{\vecL}{\text{vecl}}

\newcommand{\mCfl}{{\mC_{f}}_l}
\newcommand{\mCgl}{{\mC_{g}}_l}

\newcommand{\Ch}{\hat{C}}
\newcommand{\Bt}{\tilde{B}}
\newcommand{\Et}{\tilde{E}}
\newcommand{\bld}[1]{\mathrm{#1}}
\newcommand{\mLambda}{\bm{\Lambda}}
\newcommand{\mA}{\bm{A}}
\newcommand{\mB}{\bm{B}}
\newcommand{\mBp}{\mB_n}

\newcommand{\mC}{\bm{C}}
\newcommand{\mD}{\bm{D}}
\newcommand{\mE}{\bm{E}}
\newcommand{\mF}{\bm{F}}
\newcommand{\mg}{\bm{g}}

\newcommand{\mQ}{\bm{Q}}
\newcommand{\mU}{\bm{U}}
\newcommand{\mX}{\bm{X}}
\newcommand{\mV}{\bm{V}}
\newcommand{\mP}{\bm{P}}
\newcommand{\mR}{\bm{R}}
\newcommand{\mT}{\bm{T}}
\newcommand{\mW}{\bm{W}}
\newcommand{\mI}{\bm{I}}
\newcommand{\mH}{\bm{H}}
\newcommand{\mM}{\bm{M}}
\newcommand{\mN}{\bm{N}}
\newcommand{\mMh}{\hat{\mM}}

\newcommand{\mY}{\bm{Y}}

\newcommand{\vs}{\mathbf{s}}
\newcommand{\vshat}{\hat{\mathbf{s}}}
\newcommand{\vk}{\mathbf{k}}
\renewcommand{\vr}{\mathbf{r}}

\newcommand{\vv}{\mathbf{v}}
\newcommand{\vd}{\mathbf{d}}
\newcommand{\vC}{\mathbf{C}}
\newcommand{\vT}{\mathbf{T}}
\newcommand{\vTheta}{\mathbf{\Theta}}

\newcommand{\mS}{\bm{S}}
\newcommand{\mzero}{\bm{0}}
\newcommand{\mL}{\bm{L}}
\newcommand{\btheta}{\bm{\theta}}
\newcommand{\bphi}{\bm{\psi}}
\newcommand{\vbeta}{\bm{\beta}}
\newcommand{\va}{\mathbf{a}}
\newcommand{\vX}{\mathbf{X}}
\newcommand{\vchi}{\bm{\chi}}

\newcommand{\vXh}{\hat{\vX}}
\newcommand{\vrhat}{\hat{\vr}}

\newcommand{\vS}{\mathbf{S}}
\newcommand{\vm}{\mathbf{m}}
\newcommand{\vn}{\mathbf{n}}
\newcommand{\vnhat}{\hat{\mathbf{n}}}
\newcommand{\vkhat}{\hat{\mathbf{k}}}

\newcommand{\vN}{\mathbf{N}}
\newcommand{\vXhat}{\hat{\mathbf{X}}}
\newcommand{\vxhat}{\hat{\mathbf{x}}}
\newcommand{\vb}{\mathbf{b}}
\newcommand{\vA}{\mathbf{A}}
\newcommand{\vAt}{\tilde{\mathbf{A}}}
\newcommand{\ve}{\mathbf{e}}
\newcommand{\vE}{\mathbf{E}}
\newcommand{\vB}{\mathbf{B}}
\newcommand{\vl}{\mathbf{l}}
\newcommand{\vp}{\mathbf{p}}
\newcommand{\vP}{\mathbf{P}}

\newcommand{\vXf}{\mathbf{X}_f}
\newcommand{\vEt}{\tilde{\mathbf{E}}}
\newcommand{\vBt}{\tilde{\mathbf{B}}}
\newcommand{\vEw}{\mathbf{E}_W}
\newcommand{\vBw}{\mathbf{B}_W}
\newcommand{\vx}{\mathbf{x}}
\newcommand{\vXt}{\tilde{\vX}}
\newcommand{\vXb}{\bar{\vX}}
\newcommand{\vTb}{\bar{\vT}}
\newcommand{\vTt}{\tilde{\vT}}
\newcommand{\vY}{\mathbf{Y}}
\newcommand{\vBwr}{{\vBw^{(R)}}}
\newcommand{\RW}{{W^{(R)}}}
\newcommand{\mUt}{\tilde{\mU}}
\newcommand{\mVt}{\tilde{\mV}}
\newcommand{\mDt}{\tilde{\mD}}

\newcommand{\healpix}{HEALPix}

\def\be{\begin{equation}}
\def\ee{\end{equation}}
\def\ba{\begin{eqnarray}}
\def\ea{\end{eqnarray}}
\def\nn{\nonumber}
\def\ptg{\underline{Y}}
\def\mptg{\bm{\ptg}}

\newcommand{\dmh}[1]{\textcolor{blue}{(DMH: #1)}}
\newcommand{\aml}[1]{\textcolor{red}{(AML: #1)}}

\def\dg{^{\rm o}}
\def\lp{{\cal L}}
\def\p{\vec{h}}
\def\hp{\hat{h}}
\def\vhp{\vec{\hp}}
\def\ps{h}
\def\ph{\tilde{h}}
\def\vph{\vec{\ph}}
\def\h{{\cal H}}
\def\f{{\cal F}}
\def\a{{\cal A}}
\def\d{\delta}

\def\vf{\vec{f}}
\def\mf{\bm{\f}}

\def\vh{\bm{\h}}
\def\tr{{\rm Tr}}
\def\ta{\hat{\Theta}}
\def\tf{\bar{\Theta}}
\def\tl{\tilde{\Theta}}
\def\ctta{C^{\ta \ta}}
\def\mctta{\mC^{\ta \ta}}
\def\cttu{\mC^{\Theta \Theta}}
\def\cttn{\mC^{NN}}
\def\vtf{\vec{\tf}}
\def\vtl{\vec{\tl}}
\def\vta{\vec{\ta}}
\def\vtr{\vec{\tr}}


\title{Estimators for CMB statistical anisotropy}

\author{Duncan Hanson}
\affiliation{Institute of Astronomy and Kavli Institute for Cosmology, Madingley Road, Cambridge, CB3 0HA, UK.}

\author{Antony Lewis}
\homepage{http://cosmologist.info}
\affiliation{Institute of Astronomy and Kavli Institute for Cosmology, Madingley Road, Cambridge, CB3 0HA, UK.}

\date{\today}

\begin{abstract}
We use quadratic maximum-likelihood (QML) estimators
to constrain models with Gaussian but
statistically anisotropic Cosmic Microwave Background
(CMB) fluctuations, using CMB maps with realistic sky-coverage
and instrumental noise.
This approach is optimal when the anisotropy is small, or
when checking for consistency with isotropy.
We demonstrate the power of the QML approach by applying it to
the WMAP data to constrain several models which modulate
the observed CMB fluctuations to produce a statistically anisotropic
sky.
We first constrain
 an empirically motivated spatial modulation of the observed
 CMB fluctuations, reproducing marginal evidence for a dipolar
 modulation pattern with amplitude $7\%$ at $l \alt 60$,
 but demonstrate that the effect decreases at higher multipoles and is
 $\alt 1\%$ at $l \sim 500$.
We also look for evidence of a
  direction-dependent primordial power spectrum,
  finding a very statistically significant quadrupole signal nearly aligned with the ecliptic plane; however
  we argue this anisotropy is largely contaminated by observational systematics.
Finally, we constrain the anisotropy due to a spatial modulation of adiabatic and isocurvature
primordial perturbations, and discuss the close relationship
between anisotropy and non-Gaussianity estimators.
\end{abstract}

\maketitle

\pagenumbering{arabic}

\section{Introduction}
The temperature fluctuations of the Cosmic Microwave Background (CMB)
are often
assumed to be a realization of a statistically isotropic Gaussian random
field. Decomposing the temperature fluctuations $\Theta(\Omega)$ into
harmonic coefficients
\be
\Theta_{lm} = \int \ud\Omega Y_{lm}^*(\Omega) \Theta(\Omega),
\ee
the covariance of the CMB is then given by
$C_{lm,l'm'}
= \langle \Theta_{lm} \Theta^{*}_{l'm'} \rangle
= \delta_{ll'} \delta_{mm'} C_{l}^{\Theta \Theta}$,
and the statistical properties of the fluctuations are completely described
by the power spectrum $C^{\Theta\Theta}_{l}$. The assumption of statistical isotropy
is well motivated theoretically, both as an application of the Copernican principle
and as a prediction of more detailed cosmological models such as inflation.
It is a central tenet of modern precision cosmology and the standard $\Lambda$CDM
model, and as such needs to be rigorously tested.
There are already tantalizing hints in the WMAP data for violation of
statistical isotropy, for example alignments of low multipoles~\cite{Tegmark:2003ve,Bielewicz:2004en,Copi:2005ff},
the axis of evil~\cite{Land:2005ad}, power asymmetries~\cite{Eriksen:2003db,Hansen:2008ym},
and the cold spot~\cite{Cruz:2006fy}. Many of these oddities
have been discovered in the absence of a proposed model, and so the
degree of \textit{a posteriori} bias which they are subject to is
difficult to assess.

In this paper, we will discuss estimators which can be used
to constrain the parameters of many CMB models which are Gaussian but
perturbatively anisotropic, such that the covariance
contains small off-diagonal elements.
There are many good reasons to consider such models.
Secondary effects that are linear in the CMB temperature result in
a guaranteed signal of this form at the $\clo(10^{-3})$ level:
gravitational lensing~\cite{Lewis:2006fu}, patchy reionization \cite{Dvorkin:2008tf}
and the doppler shifting due to the motion of our frame relative to the CMB,
for example, may be thought of in this context and should become observable with the
upcoming generation of CMB measurements.
Many non-linear effects and non-standard models
can be considered as a fixed modulation of an initially Gaussian
field, which is still Gaussian but anisotropic if the modulation is
considered fixed.
More speculatively, recently
proposed anisotropic models of inflation could lead to primordial fluctuations that are
anisotropic~\cite{Gumrukcuoglu:2007bx,Ackerman:2007nb}.
A non-zero primordial bispectrum can also be considered in this framework,
where there is a power anisotropy correlated to the temperature, although the current data indicates
that any non-Gaussianity consistent with the expectations from inflationary models is small.
Models of modulated pre-heating~\cite{Bond:2009xx}, inflationary
bubble collisions (e.g.~\cite{Chang:2008gj}), topological defects (e.g.~\cite{Cruz:2008sb}),
and various other interesting scenarios could also give observable
effects that may show up in simple anisotropy estimators motivated by Gaussian anisotropic models,
although the signature would typically not be Gaussian in these cases.
In the case that the CMB is actually isotropic, the estimators which
we will discuss simply place optimal, minimum-variance constraints on the degree of anisotropy, and hence provide
valuable tests of consistency and systematics.

Assuming that the instrumental noise is also Gaussian,
the rigorous approach to an analysis of
these models is clear: calculate
the log-likelihood ${\cal L}$ of the observed CMB, given by
\be
-\lp(\vta | \p) = \frac{1}{2} \vta^{\dag} (\mctta)^{-1} \vta + \frac{1}{2} \ln \det (\mctta),
\label{eq:log_likelihood}
\ee
where $\p$ are parameters characterizing the anisotropy, $\vta$ is the observed CMB
and $\mctta \equiv \cttu + \cttn$ is its covariance, incorporating the theoretical (anisotropic) covariance
as well as instrumental noise. For modern datasets with hundreds of thousands
of modes, the full covariance matrix $\mctta$ is unmanageably large. To work with
it directly therefore requires either artificial limitation of the analysis to some subset of the data
\cite{Gordon:2006ag, Hoftuft:2009rq}, or exploitable sparseness \cite{Groeneboom:2008fz}.
In this work, we will take a quadratic estimator approach, expanding the likelihood
to low order in the anisotropy. This approach is not a new one: it was originally used for the purpose of lens
reconstruction by Ref.~\cite{Hirata:2002jy}, and many of the estimators that we will discuss
here have also been derived for full-sky coverage as minimum-variance estimators with quadratic form
\cite{Dvorkin:2007jp, Dvorkin:2008tf, Pullen:2007tu, ArmendarizPicon:2005jh,Prunet:2004zy}.
In this paper we will generalize these estimators for application to
real data, and show that in special cases where it is actually feasible to perform
the more computationally intensive exact-likelihood analysis the
quadratic approach produces effectively identical results.

Although we will frame our discussion here on Gaussian but statistically anisotropic models,
we note that every statistically anisotropic Gaussian model is related to a statistically isotropic but
non-Gaussian model: if there is a preferred direction, taking the direction as being a random
variable (e.g. by a random rotation) makes the distribution statistically isotropic at the expense
of complicating the statistics. In the statistically isotropic interpretation the anisotropy estimators
we discuss would always have zero expectation, but the non-zero disconnected four-point
function gives the estimators a variance above that expected for a Gaussian isotropic field,
giving an equivalent means of detection. This interpretation is particularly useful for e.g. lens
reconstruction and patchy reionization, where the particular realization of the anisotropy,
as constrained by the likelihood estimator, is sometimes of less interest than its statistics, contained in the
power spectrum of the estimates.

We shall focus on the CMB temperature, since this is measured with much lower noise than the polarization, especially on smaller scales.
However if there is significant evidence for anisotropy, polarization information would
ultimately be an excellent way to further constrain the origin of the
signal: an anisotropic power spectrum at recombination for example, should give a
consistent signal in the polarization, but if the signal is local --
e.g. due to an unknown foreground or instrumental systematic --
the polarization signal could be quite different. The generalization of these estimators to the polarized case is in principle straightforward.

\section{Anisotropy estimators}
\label{sec:anisotropy_estimators}

We begin by introducing the methodology of anisotropy estimation as
a likelihood maximization, loosely following Hirata and Seljak who
pioneered this approach for CMB lensing \cite{Hirata:2002jy}.

Differentiating the likelihood of Eq.~(\ref{eq:log_likelihood}) with
respect to
a set of parameters $\p$ which characterize the anisotropy gives
\ba
\frac{\d \lp}{\d \p^\dag}
\colswitch{=}{&=&}
-\frac{1}{2} \vta^\dag (\mctta)^{-1} \frac{\d \mctta}{\d \p^\dag} (\mctta)^{-1}
\colswitch{}{\vta \nn \\ && }
+ \frac{1}{2} \vtr \left[(\mctta)^{-1} \frac{\d \mctta}{\d \p^\dag} \right].
\ea
The trace term results in a ``mean field'' over realizations of the observed CMB.
To see this, consider the identity
 $\tr(\mA) = \langle \vx^\dag \mA \mC^{-1} \vx \rangle$, where $\mA$
is any matrix and $\vx$ is a vector
of Gaussian random variables with covariance $\mC$.
Making this substitution with $\mC = \mctta$
and maximizing the likelihood by setting $\d \lp / \d \p^\dag = 0$ gives the simple equation
\be
\frac{\d \lp}{\d \p^\dag}  = \left< \vh \right>  - \vh= 0,
\ee
where
\be
\vh = \frac{1}{2} \left[ (\mctta)^{-1} \vta \right]^{\dag} \frac{\d \mctta}{\d \p^\dag} \left[ (\mctta)^{-1} \vta \right].
\label{eq:H}
\ee
The maximum-likelihood (M-L) point can be determined iteratively using Newton's method
\be
\p_{i+1} = \p_{i} - \left[ \left. \frac{\d}{\d \p^\dag}(\left< \vh \right> - \vh)^\dag \right|^\dag_i \right]^{-1}( \left< \vh \right>_i  - \vh_i),
\label{eqn:ml_iter}
\ee
where quantities subscripted with $i$ are evaluated for the estimate
$\p_i$ of the $i^{\rm th}$ iteration.
Note that Eq.~\eqref{eqn:ml_iter} is a very general way to maximize a
Gaussian likelihood when the covariance is perturbatively linear in a set of parameters.
It leads to the power spectrum
estimator of Ref.~\cite{Tegmark:1996qt}, for example, if one takes the $\p$ parameters
to encode the CMB power spectrum.
We shall assume that the noise and normal cosmological parameters are known,
so we are only maximizing the anisotropy parameters; if the assumed parameters
(and hence isotropic power spectrum) are incorrect this could lead to
small biases in the estimator.


We are working under the assumption that any anisotropy which we will
be studying is ``weak'', and so a single iteration of Eq.~(\ref{eqn:ml_iter}), starting from
$\p=\vec{0}$ should give a sufficiently accurate estimate of $\vec{h}$.
For simplicity, the derivative term is replaced with its ensemble average
\ba
\left< \frac{\d}{\d \p^\dag}(\left< \vh \right> - \vh)^\dag \right>\nn &=&
\left< \frac{\d}{\d \p^\dag} \frac{\d \lp}{\d \p}  \right>  \nn \\
&=& \left< \frac{\d \lp}{\d \p^\dag} \frac{\d \lp}{\d \p}  \right> \nn\\
&=&\left[\left< \vh \vh^{\dag} \right> - \left< \vh \right> \left< \vh \right>^{\dag}\right]\nn\\
&=& \mf,
\ea
where the equality between the second and third lines is dictated by
the normalization of the likelihood and $\mf$ is the Fisher matrix,
considering the likelihood to be a function of the parameters $\p$.
Putting all of this together, we have an approximate, quadratic
maximum-likelihood (QML) estimator of the form
\be
\vhp = \mf^{-1} [\vph - \langle \vph \rangle].
\ee
The inverse Fisher matrix can be thought of as the estimator
normalization, as well as its covariance
in the limit of no anisotropy. The quadratic part of the estimator is given by
\ba
\vph = \vh_0 \colswitch{=}{&=&}
\frac{1}{2} \vtf^\dag \frac{\d \mctta}{\d \p^\dag}\vtf
\colswitch{}{\nn \\}
\colswitch{=}{&=&}
 \frac{1}{2} \sum_{lm, l'm'} \left[\frac{\d \ctta _{lm,\ l'm'}}{\d \p^\dag} \right] \tf^{*}_{lm} \tf_{l'm'},
\label{eqn:hest}
\ea
where $\vtf = (\mctta)^{-1}|_{0} \vta$ is the observed sky after
application of inverse-variance filtering
with $\p = \vec{0}$. In the limit of weak anisotropy
this QML estimator saturates the Cramer-Rao inequality,
and so is optimal in the minimum-variance sense. In practice, ``weak''
means non-detection, and so this form of quadratic estimator
is excellent for testing statistical isotropy, but
needs to be treated with care if a significant detection is made
(as is soon expected to be the case, for example, with CMB lensing).
In this work we have occasion to compare some of our results to exact
likelihood calculations by other authors, and find good agreement.

Thus, the QML formalism makes it straightforward to construct
estimators for any form of Gaussian but statistically anisotropic
model which is accurately parameterized as a linear function of
a set of parameters $\p$.
To assess the significance of a possible detection,
a useful statistic is the $\chi^2$ value of the estimate, given by
\be
\chi^2(\hat{\vf}) = \hat{\vf}^{\dag} \mf \hat{\vf}.
\label{eqn:chi2f}
\ee
For $n$ constrained parameters, the measured value can be compared
to the cumulative distribution function (CDF) of the $\chi^2_{n}$
distribution to produce an intuitive ``p-value''
figure of merit, which for this paper we will define to be the
probability of attaining a greater value of $\chi^2$ in an
isotropic model.

There are several advantages of the quadratic approach:
\begin{list}{}{\leftmargin=0.0em}
\item{\textit{Speed:}} Provided that an inverse variance filtered
  sky-map has already been calculated, the evaluation of the $\ph$
  terms is at most ${\cal O}(l_{\rm max}^4)$. In practice, the QML
  approach often results in estimators which have a fast
  implementation in real space, making them ${\cal O}(l_{\rm max}^3)$.
  Simulations are often required to determine the mean-field and
  estimator normalization, but the number of these is generally
  less than are needed for an exploration of the likelihood.

  In many situations, QML estimators also enable the study
  of a much larger model space than could conceivably be explored with
  exact likelihood evaluations. For example, the modulation anisotropy models
  which we will use in the following sections result in QML estimators
  which produce an entire modulation sky-map, rather than just
  estimating a small number of its coefficients as has been done in
  previous exact likelihood analyses. This can help to
  identify the source of contaminating signals.
\item{\textit{Systematic and jacknife tests:}}
The quadratic estimator
 approach is amenable to certain systematic and jacknife tests for
 which there are no analogues in an exact likelihood exploration.
 The effects of correlated noise, for example, can be avoided
 straightforwardly by forming the anisotropy estimates from
 cross-correlations between maps with different noise realizations.
 In this case, the likelihood approach which was used to derive our
 estimator is invalid. In particular, the estimator normalization
 and the reconstruction covariance are no longer equal.
 For full-sky coverage and homogeneous noise they
 may be derived using a minimum-variance approach,
 similar to that which is used in Ref.~\cite{Okamoto:2003zw}.
 For the purpose of significance testing however, it is irrelevant
whether the estimates are properly normalized, only that their
covariance is accurately characterized. The $\chi^2$ value defined
in Eq.~\eqref{eqn:chi2f}, for example, is still a valid measure of
significance, even though the inverse Fisher matrix no longer properly normalizes
the estimates.

The quadratic approach also enables the use of simple jacknife tests
in harmonic space.
The sum of Eq.~\eqref{eqn:hest}, for example, can be broken up
into contributions from different multipole ranges at no additional
computational cost.
Provided that the inverse variance filter does not significantly mix
the modes of the observed CMB across the chosen multipole ranges,
this can be used to construct anisotropy estimators in bands, and
localize the origin of any detected anisotropy in harmonic space.

In the quadratic approach the treatment of additional Gaussian but
anisotropic systematics is also clear. Provided that such effects
can be incorporated into the simulation and inverse-variance filtering
steps they will be subtracted from the anisotropy estimates and their additional
contribution to the variance of the estimates will be included in
$\f^{-1}$.
The effect of beam asymmetries and sidelobe pickup, for example,
fall neatly into this category,
though a method for correcting the contaminating signals could be better in practice.

\end{list}

We now proceed to apply the QML formalism to several anisotropic
modulation models, which we constrain using the WMAP data.

\section{Implementation}
We will test the assumption of statistical anisotropy in the WMAP
5-year maps, which are provided in \healpix\ format
at $\nside=512$~\cite{Hinshaw:2008kr}.
Specifically, we will study the Q-, V- and W-band data
at $40$, $60$ and $90$Ghz.
We limit our analysis of the
Q-band data to $l_{\rm max}=300$, as its power spectrum becomes
clearly contaminated by point sources at higher multipoles, which we
do not attempt to model.
The mean field and estimator normalization are generally determined for each data map using
 Monte-Carlo simulations, although we will sometimes use analytic
 approximations to the normalization when this is not possible.

For the high-resolution WMAP data, the inverse-variance filtering
operation is too costly to perform directly. Instead,
we use the conjugate gradients technique to iteratively solve \cite{Oh:1998sr}
\be
[(\cttu)^{-1} + \mptg^{\dag}\mN^{-1}\mptg] \cttu \vtf = \mptg^\dag \mN^{-1}
\vta.
\label{eqn:inv_filter}
\ee
Here $\cttu$ is the intrinsic CMB covariance, diagonal in harmonic space in the
limit of no anisotropy.
Throughout this work, we will use a fixed flat $\Lambda$CDM cosmology for $\cttu$,
with standard parameters $\{ \Omega_b, \Omega_c, h, n_s, \tau, A_s \} = \{ 0.05, 0.23, 0.7, 0.96, 0.08, 2.4\times 10^{-9} \}$,
which is consistent with the WMAP5 best-fit power spectrum~\cite{Dunkley:2008ie}. The data itself is
contained in the map vector $\vta$ and $\mptg$
is the pointing matrix, which provides the mapping between a harmonic
space signal and the observed map:
\be
\ptg_{i, (lm)} = B_{l} Y_{lm}(i),
\ee
where $B_{l}$ is the transfer function for the beam, which is assumed
to be symmetric. Finally,
$\mN^{-1}$ is the noise model, which we take in real space to be
\be
\mN^{-1} = \mN^{-1}_{\rm pix} - \mN^{-1}_{\rm pix} \mT^{T} [\mT^{T} \mN^{-1}_{\rm pix} \vT]^{-1} \vT \mN_{\rm pix}^{-1}.
\ee
Here $\mN^{-1}_{\rm pix}$ is the covariance matrix of the map noise, which is taken to be
uncorrelated between pixels. To effect a sky-cut, $\mN^{-1}_{\rm pix}$ is taken to be zero for masked pixels.
$\vT$ is an $n_{\rm tmpl.} \times n_{\rm pix}$ matrix of template maps to be
projected out of the data. Unless otherwise noted, we use this only to project
out the four templates corresponding to monopole and dipole modes.

Within the conjugate gradients approach, the solution of
Eq.~(\ref{eqn:inv_filter}) requires careful preconditioning to be fast enough
for the large number of Monte-Carlo simulations which we perform. We implement
the multigrid preconditioner given by Ref.~\cite{Smith:2007rg}, which is the fastest to date.
For a fractional error of less than ${\cal O}(10^{-6})$ in each
mode of the calculated $\vtf$ field, we find that our implementation
has a typical cost of ten minutes
on a 2GHz processor, evaluated to $l_{\rm max}=1000$
for WMAP noise with a galactic cut.

\section{Results}

\subsection{Modulation on the sky}
\label{sec:modulation_on_the_sky}

A popular form of modulation anisotropy which has been tested in the literature is given by \cite{2005PhRvD..72j3002G}
\be
\Theta_f(\vnhat) = [1+f(\vnhat)]\Theta_f^{i}(\vnhat),
\label{mod_form}
\ee
where $\Theta_f^{i}(\vnhat)$ is some intrinsic statistically
isotropic CMB temperature, $f(\vnhat)$ is a modulating field, and the $f$
subscript denotes restriction to some range of angular scales (e.g. $l\le \lmax$).
The dipole part of $f(\vnhat)$ has received particular attention since the work of Ref.~\cite{Hansen:2004vq},
 which found evidence for a large-scale hemispherical power asymmetry in the WMAP data.
 This pathfinding work was followed by more rigorous likelihood analyses~\cite{Spergel:2006hy:v1,Gordon:2006ag,Eriksen:2007pc},
 and recently Refs.~\cite{Hansen:2008ym, Hoftuft:2009rq} have extended
 the analysis to smaller angular scales,
 arguing for increased detection significance as more data is
 added.
 The exact likelihood analysis is currently limited by computational requirements to multipoles $l \le 80$,
 but with the QML approximation to the maximum-likelihood estimator no such difficulties arise,
 so we can extend our analysis to the limit of current observations.

We note that the doppler effect giving the CMB kinematic dipole also results in a dipole modulation
of the assumed form, with $f(\vnhat)=\vbeta \cdot \vnhat$. Assuming a small $\clo(10^{-5})$
primordial dipole, $\vbeta$ is expected to have amplitude $0.0012$ in the direction
$(l,b)=(260\dg,50\dg)$ of the observed total dipole. This signal
should be present on all scales
along with
an aberration effect which is degenerate with a CMB lensing dipole. Although we will see that
this effect is an order of magnitude too small to be observed with WMAP, it should be detected
with high significance in the data of the recently launched Planck satellite \cite{ChallinorSollom}.
Any robust signal detected at a higher level (amplitude $\gg 0.001$) would be an indicator of cosmological
statistical anisotropy, or equivalently non-Gaussianity.

The simplest quadratic estimator for the modulation is just $\ta^2$ - an estimator of the temperature variance in each pixel.
In a statistically isotropic model a map of $\ta^2$ should on average
have only a monopole, and there should be no structure which
is not consistent with noise or CMB fluctuations. The QML estimator that
we derive essentially generalizes this to optimally account for noise and cosmic variance.

The temperature moments of Eq.~\eqref{mod_form} for the modulation anisotropy are given by
\be
\Theta_{lm} = \Theta^{i}_{lm} + \sum_{l' m',l'' m''} \Theta^{i}_{l' m'} f_{l'' m''} \int d\Omega Y^*_{lm} Y_{l'm'} Y_{l''m''},
\ee
where $l,l'$ are restricted to the range of $l$ under consideration.
To first order in $f_{lm}$, the resulting covariance is therefore
\begin{multline}
C_{l_1m_1,l_2m_2}
\equiv \la \Theta_{l_1 m_1} \Theta_{l_2 m_2}^*\ra
 \\
=\delta_{l_1 l_2} \delta_{m_1 m_2} C_{l_1}
+  \sum_{lm}
f_{lm} [C_{l_1} + C_{l_2}]
\int d\Omega Y_{lm} Y^*_{l_1m_1} Y_{l_2m_2} .
\end{multline}
Using Eq.~\eqref{eqn:hest} results in a QML estimator constructed from
\be
\ph^{f}_{lm} =
\int d\Omega Y_{lm}^{*} \left[ \sum_{l_1 m_1}^{l_{\rm max}} \tf_{l_1 m_1} Y_{l_1 m_1} \right] \left[ \sum_{l_2 m_2}^{l_{\rm max}} C_{l_2} \tf_{l_2 m_2} Y_{l_2 m_2} \right].
\label{modulation_h}
\ee
On the full sky the expected value depends only on the off-diagonal part of the underlying
temperature realization covariance matrix, and is essentially an optimally
weighted combination of the bipolar spherical harmonic coefficients introduced in Ref.~\cite{Hajian:2003qq}.
Here we have written it in a form convenient for fast numerical evaluation in terms of real-space fields;
in harmonic space it can be written as
\ba
\ph^{f}_{lm}
\colswitch{=}{&=&}
\frac{1}{2}\sum_{l_1,m_1,l_2,m_2} (-1)^{m_1}
\sqrt{\frac{(2l+1)(2l_1+1)(2l_2+1)}{4\pi}}
\colswitch{}{\nn \\ &&}
\threejz{l}{l_1}{l_2}
\threej{l}{l_1}{l_2}{m}{-m_1}{m_2} (C_{l_1}+C_{l_2}) \tf_{l_1 m_1}
\tf^*_{l_2 m_2}. \nn
\ea
Due to the triangle constraint $|l_1-l_2| \le l\le |l_1+l_2|$ from the 3-jm symbols, and the constraint that
$l+l_1+l_2$ is even, we see that the dipole part of $\ph^{f}_{lm}$ is due to the off-diagonal $l_1,l_1+1$
part of the covariance, and similarly the quadrupole part depends on the $l_1,l_1$, and $l_1,l_1+2$ correlations.
Note that the estimator is independent of the modulation pattern, amplitude and orientation, so the estimator
only needs to be calculated once to recover all multipoles --- there
is no need to marginalize over all the possible orientations.

The Fisher matrix is determined by the trispectrum of the inverse-variance filtered CMB.
The CMB is assumed to be Gaussian
and so this consists only of disconnected terms.
A useful tool for forecasting is the portion of the Fisher matrix due to the isotropic terms,
i.e. those in which any coupling between modes of the
observed CMB are taken to be zero
(i.e. we ignore the non-diagonal elements of $\mctta$ so $\ctta_{ll'} \rightarrow \delta_{ll'}\Ctot_l$). This is given by
\begin{multline}
\left[ \f^{ff}_{\rm iso}  \right]_{l m, l' m'} = \delta_{ll'} \delta_{mm'}
 \\
\times\sum_{l_1,l_2} \frac{(2l_1+1)(2l_2+1)}{8\pi}
\threejz{l}{l_1}{l_2} ^2 \frac{(C_{l_1}+C_{l_2})^2}{\Ctot_{l_1} \Ctot_{l_2}}.
\end{multline}
The corresponding estimator noise we denote as $N_{l}^{ff} = [ \f^{ff}_{\rm iso} ]^{-1}_{l l}$.
In practice, for small sky-cuts and roughly homogeneous uncorrelated instrumental noise
this is quite a good approximation to the true covariance.
For full-sky coverage and homogeneous noise, it is exact.
For a cosmic-variance limited reconstruction, $N^{ff}_{l}$ scales with
the number of observed modes, as $l_{\rm max}^{-2}$ at
high-$l$.
For reconstruction of the modulation dipole $f_{1m}$, for example, we find $N^{ff}_{1} \sim 6.24 l_{\rm max}^{-2}$.
\begin{figure}
\begin{center}
\includegraphics[width=\columnwidth]{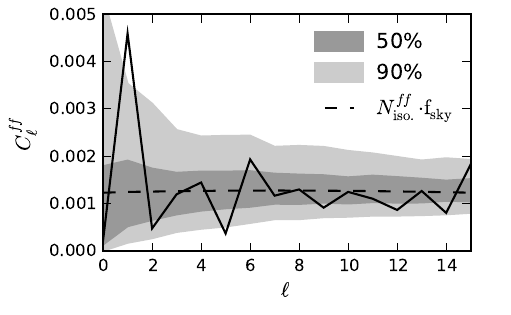}
\end{center}
\caption{Pseudo-$C_{l}$ of the $f_{l m}$ reconstruction for the WMAP V-band foreground-reduced data, with KQ85 mask and $l_{\rm max}=64$ (black solid). The full-sky, isotropic normalization was used rather than the actual inverse Fisher matrix. The $[25,75]\%$ (dark gray) and $[5,95]\%$ (light gray) confidence intervals measured from Monte-Carlo simulations are overlaid.
}
\label{fig:modulation_lmax64_pseudo_cl}
\end{figure}

\begin{figure}
\begin{center}
\includegraphics[width=\columnwidth]{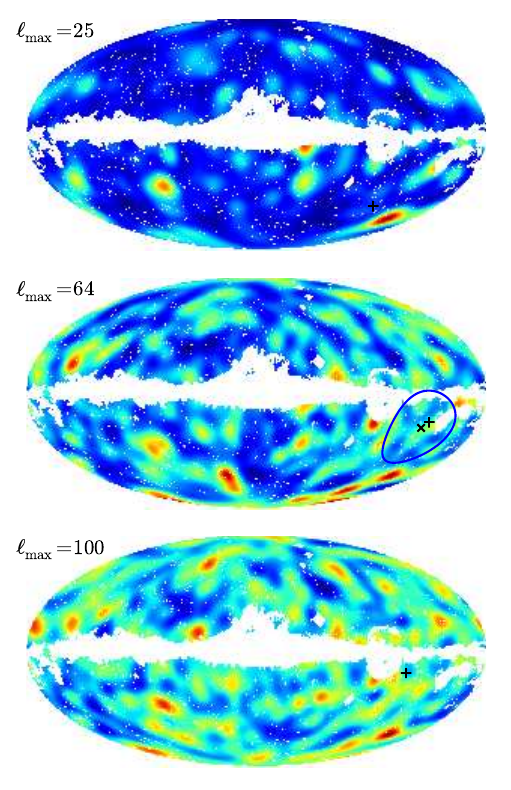}
\end{center}
\caption{Reconstructed maps of $f(\vnhat)$ for three values of $l_{\rm
    max}$, smoothed with a ten degree beam.
  We have used the isotropic normalization for simplicity,
  which is invalid close to the sky-cut, but works well otherwise.
  The `$+$' symbols mark the peak of the QML dipole. The `$\times$' symbol
  and ring in the $l_{\rm max}=64$ plot marks the
M-L dipole and error
found by \cite{Hoftuft:2009rq}, which agrees well with our result.}
\label{fig:modulation_maps_lmax_25_64_100}
\end{figure}

In Fig.~\ref{fig:modulation_lmax64_pseudo_cl} we show the pseudo-$C_{l}$
power spectrum of a typical $f_{l m}$ reconstruction.
For simplicity we use the isotropic normalization for this plot,
with confidence intervals determined by Monte-Carlo simulation.
The reconstruction variance compares very well to the full-sky analytical result,
scaled by the unmasked sky-fraction.
In Fig.~\ref{fig:modulation_maps_lmax_25_64_100} we show maps of the
reconstructed $f(\vnhat)$ for $l_{\rm max} = 25,\ 64,\ 100$, smoothed with a ten degree beam.
The modulation reconstruction brings the hot and cold spots of the observed CMB into sharp relief.
The CMB cold spot \cite{Cruz:2004ce}, for example, is distinctly visible in all three maps.

\begin{figure}
\begin{center}
\includegraphics[width=\columnwidth]{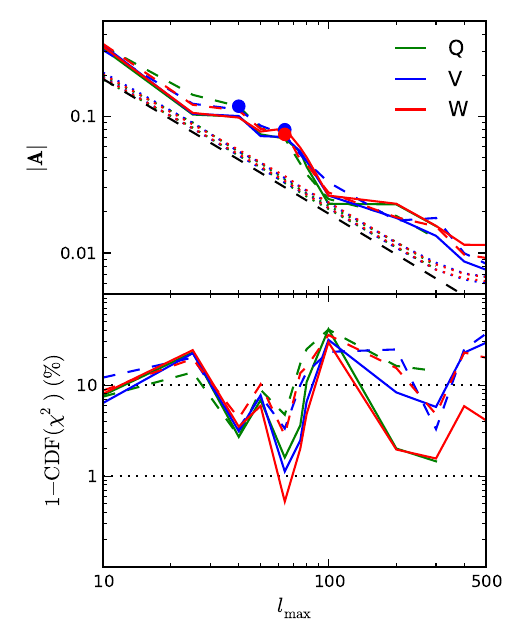}
\end{center}
\caption{
Summary of modulation dipole results for the foreground-reduced WMAP data.
Solid lines correspond to KQ85 masking, and dashed lines use the KQ75 mask.
\panel{Upper panel} Dipole
amplitudes $|\vec{A}|$ as a function of the maximum multipole used in the reconstruction. The black dashed
line gives the expected value for a cosmic-variance limited experiment, which is non-zero due to the estimator noise.
The dotted lines give the reconstruction noise spectra measured from the simulations.
They separate into two groups for KQ85 and KQ75 masking
and are well described as $f_{\rm sky}^{-1}$ times the ideal result for $l_{\rm max}<300$,
but decrease more slowly at higher-$l$ as the instrumental noise becomes non-negligible.
\panel{Lower panel} $\chi^2$
significances of the reconstructions in the isotropic model.}
\label{fig:modulation_ml_QVW_std_AS}
\end{figure}

It can be seen from Fig.~\ref{fig:modulation_lmax64_pseudo_cl} that the amplitude of
the reconstructed modulation is generally in good agreement with that expected from
our $\Lambda$CDM simulations, except for the dipole which is slightly high.
To make connection with previous work, we will now focus exclusively
on the dipole modulation,
restricting our reconstruction to the three modes of the dipole,
for which we determine the full Fisher matrix by Monte-Carlo.
The three dipole modes may be simply related to the amplitude and galactic coordinates
of a dot product modulation $f(\vnhat) = \vA \cdot \vnhat$.
In the upper panel of Fig.~\ref{fig:modulation_ml_QVW_std_AS} we show estimates of this
modulation amplitude $|\vA|$ for the WMAP data, as well as the expectation value
(due to the estimator noise) for a full-sky cosmic variance limited
measurement.
The observed amplitudes generally
lie above the full-sky expectation. This is partially due to the presence of instrumental noise
and a sky-cut, particularly at high-$l$ where the WMAP data becomes noise dominated.
In the lower panel of Fig.~\ref{fig:modulation_ml_QVW_std_AS} we plot
the analytical significance
levels of the $\chi^2$ values calculated following Eq.~\eqref{eqn:chi2f}, given three degrees of freedom.
We can see that all of the observed p-values are somewhat low, typically at the five percent significance level.
Interestingly, the probability of the observed $f_{1m}$ in an isotropic model is smallest at $l_{\rm max} = 40,\ 64$,
which are the two most previously studied values.

We note that in their region of overlap, our measurements agree well with those of Ref.~\cite{Hoftuft:2009rq}.
Our measured $(A,l,b)$ values for  $l_{\rm max}<100$ are all within error of those which they quote, and consistent with a preferred
direction of $(l,b)=(225\dg,-22\dg)\pm24\dg$.
Their quoted significance values also agree well with the square-root of our $\chi^2$ values,
although we note that for the purposes of testing $\Lambda$CDM, they
should be treated as $\chi^2_3$ significances, rather than Gaussian ones.
Fig.~\ref{fig:modulation_ml_QVW_std_AS} shows that the large
modulation indicated by the low-$l$ data does not persist on smaller scales however,
consistent with the tight constraint on the anisotropy in the quasar distribution~\cite{Hirata:2009ar}.
\begin{figure}
\begin{center}
\includegraphics[width=\columnwidth]{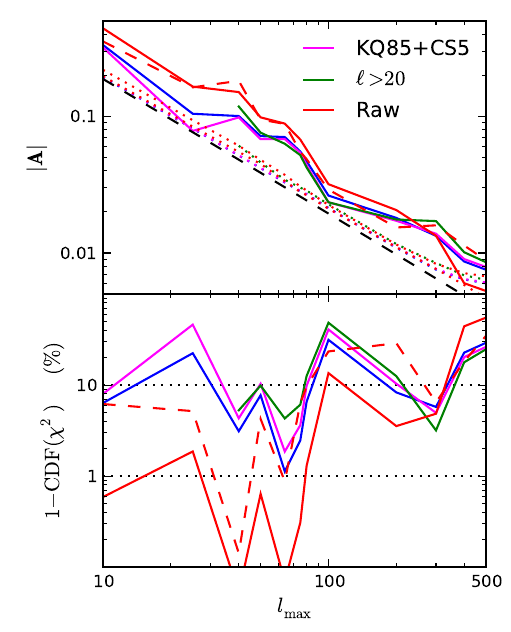}
\end{center}
\caption{
Sensitivity of modulation results to various tests, similar to Fig.~\ref{fig:modulation_ml_QVW_std_AS}.
All data are for WMAP V-band. Blue solid is the V-band foreground-reduced data, with KQ85 mask.
Magenta is with the KQ85+CS5 mask. Green is for reconstructions with $l_{\rm min}=20$.
Red lines are for raw maps, without template cleaning (solid/dashed correspond to KQ85/KQ75 masking respectively.) }
\label{fig:modulation_ml_QVW_tests_AS}
\end{figure}
There is still some tension between these measurements of $f_{1 m}$ and an isotropic model.
Although chance is a possible explanation, it is intriguing that the observed amplitude of dipole modulation
is consistently high across a large range of multipole values (although it must be remembered that the
measurement of $\vec{A}$ is cumulative, and so the estimators at different $l_{\rm max}$ can be strongly correlated).
To be rigorous, we consider here several other possible explanations for this tension:
\begin{list}{}{\leftmargin=0em}
\item{\textit{The cold spot:}}
In Fig.~\ref{fig:modulation_maps_lmax_25_64_100} we have seen that the CMB ``Cold Spot'' \cite{Cruz:2004ce}
constitutes a prominent feature in the reconstructed modulation field. It is also close to the dipole of the reconstruction.
It is possible that the large dipole amplitude is simply another detection of the Cold Spot. To test this, we perform the
modulation reconstruction with a new mask which we call KQ85+CS5, created by augmenting the KQ85 mask with a
circular cut of radius five degrees centered at $(l,b) = (208\dg,-56\dg)$.
The results of this reanalysis are given in Fig.~\ref{fig:modulation_ml_QVW_tests_AS}.
The removal of the Cold Spot has a large effect for the $l_{\rm max}=25$ reconstruction, but does not significantly
effect the reconstructed modulation at higher multipoles.
\item{\textit{Residual low-$l$ anomalies:}}
The modulation model may be motivated as a way to consistently
treat the observed low-$l$ power deficit, asymmetry, and alignment issues \cite{Spergel:2006hy:v1}.
To what extent do the low-$l$ anomalies imprint upon our measurements of $f_{l m}$ for larger values of $l_{\rm max}$?
We test this straightforwardly by excluding all of the filtered multipoles below some $l_{\rm min}$ from our analysis.
We choose $l_{\rm min}=20$, which ensures that any mixing of the anomalies at $l < 10$ due to the sky-cut will
have negligible effects on the reconstruction.
This is shown in Fig.~\ref{fig:modulation_ml_QVW_tests_AS}.
Again, this excision does not result in any significant reduction of the measured modulation.

\item{\textit{Foregrounds:}}
The consistency of the modulation amplitudes with frequency band and sky-cut makes a
foreground explanation for the observed tension seem unlikely.
We also confirm that our results are not strongly dependent on the method of foreground subtraction,
by marginalizing over the WMAP foreground templates \cite{Gold:2008kp} in the inverse-variance filtering operation,
similar to \cite{Gordon:2006ag}. Further investigation raises two interesting points, however.
To understand the order-of-magnitude for potential foreground effects,
we measure the modulation amplitudes for raw WMAP data, without foreground template subtraction.
This is plotted in Fig.~\ref{fig:modulation_ml_QVW_tests_AS}.
We can see that complete neglect of foregrounds has a large effect on the estimated modulation.
Interestingly, this increase is not strongly dependent on choice of sky-cut,
indicating that consistency under changes in the mask size
does not represent
strong evidence against foreground contamination.
Additionally interesting is that the modulation dipole induced by foregrounds is closely
aligned with the dipole determined for cleaned maps. In the quadratic estimator approach,
an additive template which is uncorrelated with the primary CMB has an additive effect on the reconstructed modulation.
The V-band template model, for example, gives a contribution to the dipole of an $l_{\rm max}=40$
analysis in the direction $(l,b) = (208\dg, -10\dg)$, due in roughly equal parts to the contribution from the
$H\alpha$ and Finkbeiner dust templates. All of these results are of course for the current foreground model,
which is projected out of the data. It seems possible, however, that residual foreground contributions which
are morphologically similar may source the tension between the modulation estimates and an isotropic model.
\end{list}

\subsection{Primordial Power Anisotropy}


\begin{figure}
\begin{center}
\epsfig{figure=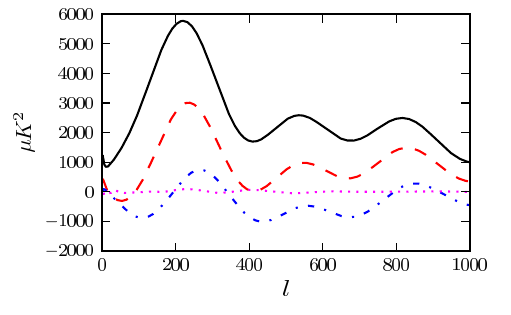,width=8truecm}
\caption{
$\sqrt{l(l+1)l_2(l_2+1)}C_{l l _2}/2\pi$ for $a(k)=1$,
with $l_2=l$ (thick solid), $l_2=l+2$ (dashed), $l_2=l+4$ (dot-dashed) and $l_2=l+20$ (dotted).
}
\label{ClPlot}
\end{center}
\end{figure}

Refs.~\cite{Ackerman:2007nb,Gumrukcuoglu:2007bx,Groeneboom:2008fz} consider
anisotropic models of the early universe, where at late times the universe isotropizes
so that the only evidence is an angular-dependent power spectrum on large scales
(but isotropic transfer functions).
The isotropic real adiabatic transfer functions $\Delta_l(k)$ are defined so that
\be
\Theta_{lm} = 4\pi i^l\int\frac{\ud^3\vk}{(2\pi)^3} \Delta_l(k) \chi_0(\vk) Y_{lm}^*(\hat{\vk})
\label{TlmTransfer}
\ee
where $\chi_0(\vk)$ is the (statistically anisotropic) primordial perturbation with power spectrum defined by
\ba
\la  \chi_0(\vk)\chi_0^*(\vk')\ra
\colswitch{=}{&=&}  (2\pi)^3 \delta(\vk-\vk') P_\chi(\vk)
\colswitch{=}{\nn \\ &=&}
(2\pi)^3 \delta(\vk-\vk') \frac{2\pi^2}{k^3} \clp_\chi(\vk).
\ea
The covariance is then given by~\cite{Gumrukcuoglu:2007bx}
\begin{multline}
C_{l_1m_1 l_2 m_2}=\\
i^{l_1-l_2} \frac{\pi}{2} \int \ud^3 \vk P_\chi(\vk)
\Delta_{l_1}(k) \Delta_{l_2}(k)
Y_{l_1 m_1}^* (\hat{\vk}) Y_{l_2 m_2}(\hat{\vk}).
\end{multline}
The anisotropy is contained in the direction-dependent $P_\chi(\vk)$, and we can construct
a QML estimator for the angular dependence of the power spectrum at a given $k$ from Eq.~\eqref{eqn:hest} using
\ba
\ph^P_{lm}(k)
\colswitch{=}{&=&}
\frac{\pi}{2}\int \ud\Omega Y_{l m}^{*} \left[ \sum_{l_1 m_1} i^{l_1}\Delta_{l_1}(k)\tf_{l_1m_1} Y_{l_1m_1} \right]
\colswitch{}{\nn \\ && \quad\times}
\left[ \sum_{l_2 m_2} i^{-l_2}\Delta_{l_2}(k)\tf_{l_2m_2} Y_{l_2m_2} \right].
\label{eqn:Pkgeneral}
\ea
Refs.~\cite{Ackerman:2007nb,Groeneboom:2008fz,Yokoyama:2008xw,Dimopoulos:2008yv} consider the particularly simple case
where the $k$-dependence is in a known function $a(k)$ so that
\be
\clp_\chi(\vk) = \clp_\chi(k)[1 + a(k) g(\hat{\vk})],
\ee
and $g(\vk) = g(-\vk)$ so $g_{lm}$ has only even-$l$ modes. The covariance in this case is
\begin{multline}
C_{l_1m_1 l_2 m_2}=
 \delta_{l_1 l_2}\delta_{m_1 m_2}C_{l_1}
\\+  \sum_{lm} i^{l_1-l_2}g_{lm}
\int \ud\Omega_\vk C_{l_1 l_2}
Y_{l m} \colswitch{(\hat{\vk})}{}
Y_{l_1 m_1}^* \colswitch{(\hat{\vk})}{}
Y_{l_2 m_2} \colswitch{(\hat{\vk})}{}
\end{multline}
where
\be
C_{l_1 l_2} \equiv 4\pi\int \ud\ln k \clp_\chi(k) a(k) \Delta_{l_1}(k) \Delta_{l_2}(k).
\label{Cldef}
\ee
The QML estimator for $g$ is therefore constructed from
\ba
\ph^g_{lm}
\colswitch{=}{&=&}
\frac{1}{2} \int \ud\Omega Y_{l m}^{*} \sum_{l_1 l_2} i^{l_1-l_2}C_{l_1 l_2}
\colswitch{}{ \nn \\&&\quad\times}
\left[ \sum_{m_1} \tf_{l_1m_1} Y_{l_1m_1} \right]
\left[ \sum_{m_2} \tf_{l_2m_2} Y_{l_2m_2} \right].
\label{eqn:primordial_estimator}
\ea
The $i^{l_1 - l_2}$ term forces the reconstruction to zero for odd-$l$.
Similar quadratic estimators have been derived before on the full sky~\cite{ArmendarizPicon:2005jh,Pullen:2007tu}.
For a pixelization such as \healpix~ which consists of
${\cal O} (l_{\rm max})$ isolatitude rings,
the azimuthal sums may be evaluated using FFTs in ${\cal O}(l_{\rm max}^{2} \log l_{\rm max})$.
 To estimate $g_{lm}$ to some maximum multipole $\Delta_l$,
the $C_{l_1 l_2}$ matrix may be treated as band-diagonal with
$\Delta_l$ bands, due to the triangle constraint on the integral
of three spherical harmonics.
This leads to a total cost of ${\cal O}(\Delta_{l} l_{\rm max}^3 \log l_{\rm max})$.
An alternative approach is to discretize the integral of Eq.~(\ref{Cldef}), in which case $\hat{g}$
may be evaluated as $n_k$ spherical harmonic transforms, where $n_k$ is the number of quadrature
points for the integral, leading to a cost of ${\cal O}(n_k l_{\rm max}^3)$.
A similar technique was applied in Ref.~\cite{Smith:2006ud}.

\begin{figure}
\begin{center}
\includegraphics[width=\columnwidth]{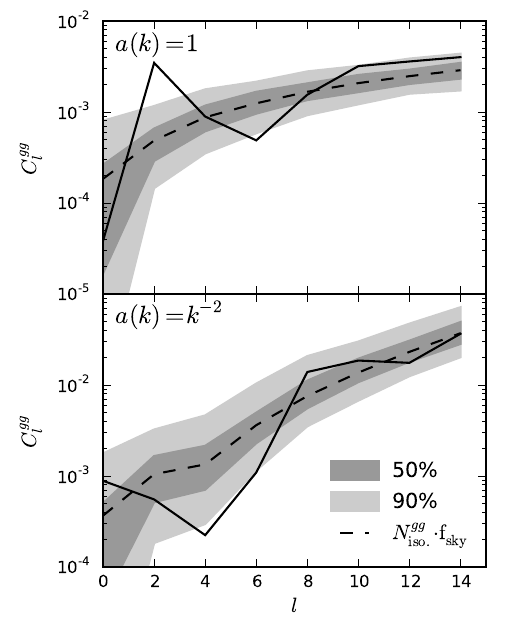}
\end{center}
\caption{
Pseudo-$C_{l}$ of the $g_{l m}$ reconstruction for the WMAP V-band foreground-reduced data,
with KQ85 mask and $l_{\rm max}=400$ (black solid).
The full-sky, isotropic normalization was used.
The $[25,75]\%$ (dark gray) and $[5,95]\%$ (light gray) confidence intervals measured from Monte-Carlo simulations are overlaid.
}
\label{fig:primordial_V_lmax400_pseudo_cl}
\end{figure}

The disconnected, isotropic component of the Fisher matrix in this case is given by
\begin{multline}
\left[ \f_{\rm iso}^{gg} \right]_{l m, l' m'}=
\delta_{l l'}\delta_{m m'} \\
\times\sum_{l_1,l_2} \frac{(2l_1+1)(2l_2+1)}{8\pi}
\threejz{l}{l_1}{l_2} ^2 \frac{C_{l_1l _2}^2}{\Ctot_{l_1}\Ctot_{l_2}}.
\end{multline}
Again we take $N_{l}^{gg} = [ \f^{gg}_{\rm iso} ]^{-1}_{ll}$. In Fig.~\ref{fig:primordial_V_lmax400_pseudo_cl} we
show the pseudo-$C_{l}$ power spectra for the reconstructed $g_{lm}$,
for $a(k)=1$ and $a(k)=k^{-2}$.
The $a(k)=k^{-2}$ results are generally in good agreement with Monte-Carlo expectations.
The $a(k)=1$ reconstruction, on the other hand, has an anomalously large quadrupole.
As with the modulation dipole of the previous section, we investigate this more thoroughly by normalizing the
estimates with the quadrupole Fisher matrix, estimated from simulations.
In Fig.~\ref{fig:primordial_lmax400_quad_map} we show the reconstructed quadrupole for the V-band data.
Groeneboom and Eriksen \cite{Groeneboom:2008fz} have performed a Bayesian analysis of the primordial
modulation quadrupole, assuming the form
$g(\vnhat)=g^{*}(\vkhat\cdot\vnhat)^2$.
They find a maximum-likelihood
quadrupole described by $g^*=0.10$ with $\vkhat$ along $(l,b)=(130\dg, 10\dg)$.
Their analysis was
based on the published version of Ref.~\cite{Ackerman:2007nb} which
neglected the $i^{l_1 - l_2}$ prefactor in the covariance;
our analysis incorporates two additional degrees of freedom and
includes the correct prefactor. The results are nonetheless quite
similar in magnitude and direction, though we find that the preferred
direction lies closer to the ecliptic poles.
\begin{figure}
\begin{center}
\includegraphics[width=\columnwidth]{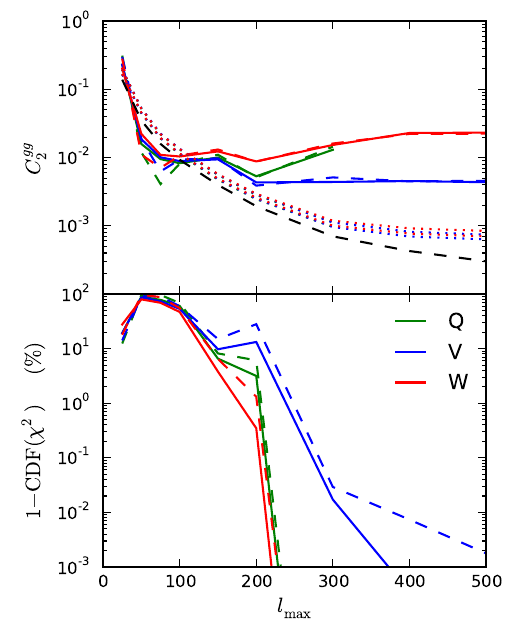}
\end{center}
\caption{
Summary of results for $g(\vn)$ quadrupole reconstructions with $a(k)=1$ for the foreground-reduced WMAP data.
Solid lines correspond to KQ85 masking, and dashed lines use the KQ75 mask.
\panel{Upper panel} Quadrupole power
as a function of the maximum multipole used in the reconstruction.
The black dashed
line gives $N_{l}^{gg}$, the reconstruction noise for a cosmic-variance limited full-sky experiment.
The dotted lines give the reconstruction noise measured from the simulations.
They separate into two groups for KQ85 and KQ75 masking
and are well described as $f_{\rm sky}^{-1} \cdot N_{l}^{gg}$ for $l_{\rm max}<300$,
but decrease more slowly at higher-$l$ as the instrumental noise becomes non-negligible.
\panel{Lower panel} $\chi^2$
significances of the reconstructions in the isotropic model.}
\label{fig:primordial_ml_QVW_std}
\end{figure}
\begin{figure}
\begin{center}
\includegraphics[width=\columnwidth]{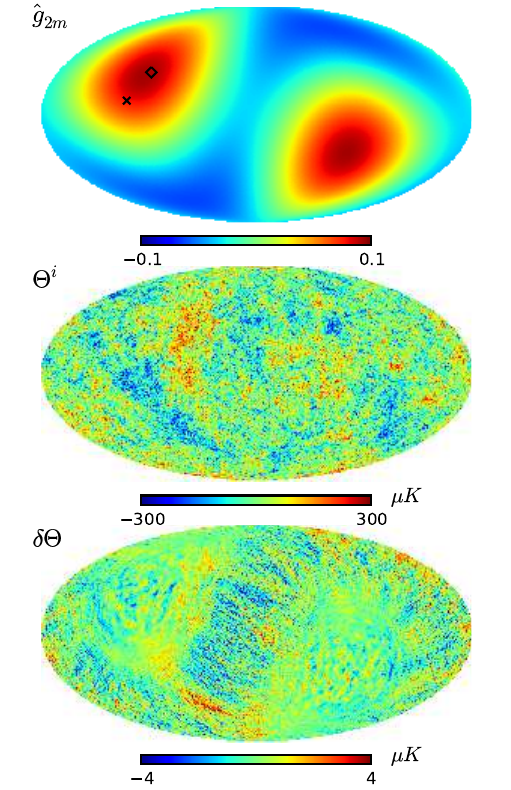}
\end{center}
\caption{
\panel{Upper panel} QML reconstruction of the $g(\hat{n})$ primordial power quadrupole, for
the WMAP V-band foreground-reduced data to $l_{\rm max}=400$.
Also shown is the preferred quadrupole direction of $(l,b)=(130\dg,10\dg)$ found
by Ref.~\cite{Groeneboom:2008fz} ($`\times'$), using incorrect $i^{l_1-l_2}$ factor) and the ecliptic north pole
at $(l,b)=(96\dg,30\dg)$ ($`\diamond'$).
\panel{Lower two panels} Isotropic and anisotropic components of a CMB
simulation with $g_{2m}$ given by the QML reconstruction, smoothed with a WMAP V-band
beam. This gives an intuitive understanding of the effects induced by
this form of anisotropy. The RMS deviations of the isotropic and
anisotropic components are $90\mu K$ and $1\mu K$ respectively.
}
\label{fig:primordial_lmax400_quad_map}
\end{figure}

The $C_{l}$ of the QML quadrupole reconstructions are given in the top panel of
Fig.~\ref{fig:primordial_ml_QVW_std} for the Q-, V- and W-band foreground-reduced datasets
with a variety of cutoff $l_{\rm max}$ in the $l_1,l_2$ sums of Eq.~\ref{eqn:primordial_estimator}.
In the lower panel are given the $\chi^2$ significances (calculated following Eq.~\ref{eqn:chi2f}),
for a distribution with five degrees of freedom. At $l_{\rm max}\alt 150$ consistent results are
seen in the Q-, V-, and W-band for all three datasets, independent of the choice of KQ75 or KQ85
masking, roughly tracing the amplitude of the estimator noise and consistent with no detection of anisotropy.
Including data from higher multipoles, a highly significant excess
power develops.
The increasing discrepancy between the V- and W-band data at $l_{\rm max}>200$
seems indicative of systematic contamination rather than a primordial
origin. To gain a better understanding of the required
systematic, we use the algorithm described in
Appendix~\ref{sec:anisotropic_simulations} to generate a CMB
realization which contains the observed modulation as the sum of
anisotropic and purely isotropic parts.
This is presented in the lower two panels of
Fig.~\ref{fig:primordial_lmax400_quad_map}.
The map of the anisotropic part gives a measure of the magnitude
and spatial dependence of the systematic required to mimic the observed signal.
The close alignment of the anomalous quadrupole modulation and
the ecliptic seen in Fig.~\ref{fig:primordial_lmax400_quad_map} suggests a
systematic associated with the WMAP scan strategy.
The two most suspicious possibilities are correlated noise and beam asymmetry effects,
which have not been accounted for in this analysis, nor included in the
Monte-Carlo simulations.

Ref.~\cite{Groeneboom:2008fz} have already partially considered the effect of
correlated noise on their results.
Working with end-to-end simulations produced by the WMAP team for the 1-year data release, they find large
effects for the W-band, although with qualitative structure different from that which is actually found in the data.
In the quadratic-estimator approach, such possible contamination can
be avoided by cross-correlating maps with different noise
realizations, as discussed in Sec.~\ref{sec:anisotropy_estimators}.
Cross-correlating data for the individual differencing assemblies (D/As) of
the V- and W-band data, we find that the anomalously large quadrupole power persists,
further indicating that correlated noise is not a likely explanation for the observed excess.

\begin{figure}
\begin{center}
\includegraphics[width=\columnwidth]{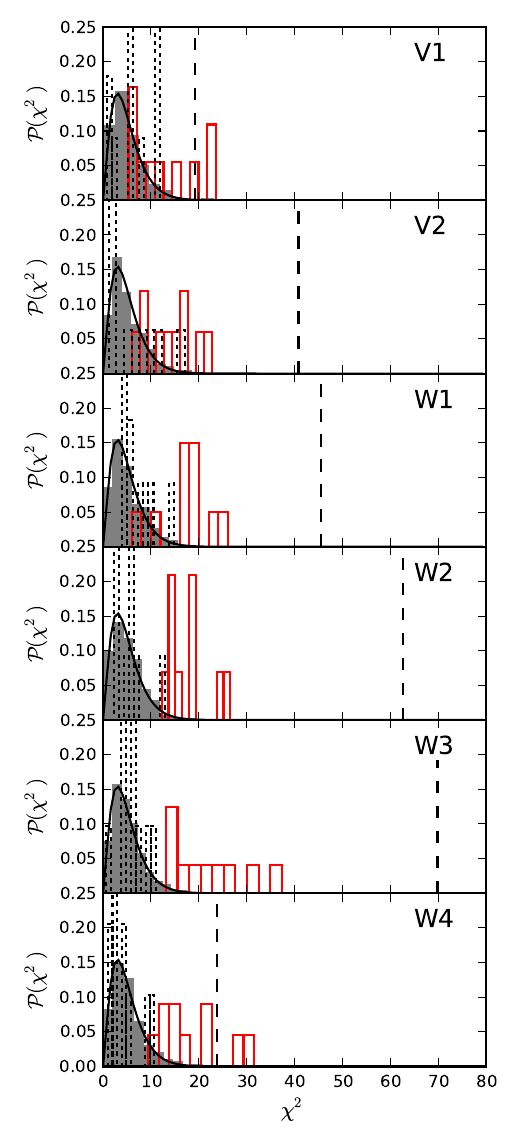}
\end{center}
\caption{Effect of beam asymmetry on the reconstruction of the
  $g(\hat{\vn})$ quadrupole with $a(k) = 1$, for the WMAP foreground-reduced data with KQ85 masking.
Solid black is the $\chi^2$ distribution with five degrees of
freedom.
Gray histograms are the distribution of values for the isotropic
simulations used to determine the estimator normalization for each D/A.
Black vertical dashed lines indicate $\chi^2$ for the actual data.
Red histograms are for the beam-convolved maps produced by Ref.~\cite{Wehus:2009zh}.
Black dotted histogram are for the input maps to the convolution,
using symmetric beam transfer functions and separate noise
realizations.}
\label{fig:primordial_wehus_quad_chi2s}
\end{figure}

The effect of beam asymmetries is more involved, however a
sample of the expected beam anisotropy effects can be found in
Fig.~11 of Ref.~\cite{Smith:2007rg} and Fig.~5 of
\cite{Wehus:2009zh}.
The $1\muK$ RMS contribution of beam asymmetries to
the observed sky is aligned with the ecliptic and
matches well with the amplitude in our
Fig.~\ref{fig:primordial_lmax400_quad_map},
making beam asymmetries a potential explanation for the observed
anisotropy.
To investigate this further, we analyze simulations of the WMAP
data released by Ref.~\cite{Wehus:2009zh}, who publish 10 maps of
simulated beam-convolved skies for each of the WMAP D/As. The distribution of the
resulting $\chi^2$ values is given in
Fig.~\ref{fig:primordial_wehus_quad_chi2s}.
The effect of asymmetric beams is seen to be large. For the V1 and W4
D/As, it completely explains the $\clo(0.1\%)$ significance of the
observed quadrupole.
We therefore conclude that
current measurements of the power modulation quadrupole are
strongly contaminated by beam asymmetry effects, which must be corrected
for to obtain true constraints on any primordial
modulation.
That the signal strongly varies between the D/As indicates either that the
simulation of  Ref.~\cite{Wehus:2009zh} are not encapsulating all of the relevant beam effects,
or that there is an additional unknown systematic. In any case the significant variations
between D/As at the same frequency provide strong evidence that the signal is systematic rather than
primordial or foregrounds; in all cases the preferred direction is close to the ecliptic.

A full analysis with beam asymmetries is beyond the scope of this paper,
however we note that in the QML approach beam asymmetry effects can simply be incorporated into the simulation
pipeline. They will then appear as a contribution to the ``mean
field'' term, and be subtracted from the reconstruction.
In principle it is necessary
to include the correlation due to beam asymmetries in the inverse
variance filter, which is too computationally expensive to perform
in general.
If the instrumental noise can be approximated as white on the
timescales which separate pixel visits, however, then the fast
algorithm presented by Ref.~\cite{Smith:2007rg}
can be used for the forward convolution operation,
which should only slow the application of the inverse variance filter
by a constant factor of ${\cal O}(20)$.
Alternatively one could attempt to correct the maps for the beam asymmetries, for example
by estimating the anisotropic contribution by forward convolutions of the observed sky,
then iteratively subtracting off the part due to beam asymmetries.


\subsection{Local primordial modulation}

\begin{figure}
\begin{center}
\epsfig{figure=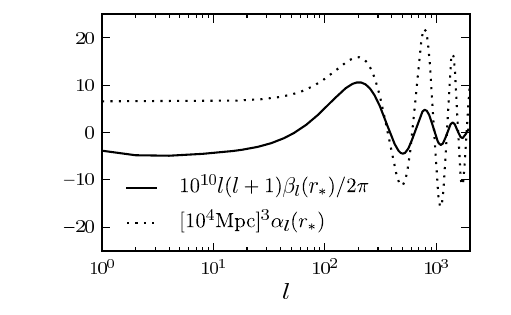,width=\columnwidth}
\caption{The $\alpha$ and $\beta$ functions for adiabatic transfer
  functions evaluated at $r_* =
  14164$Mpc, the comoving distance to the maximum visibility at the
  last scattering surface for our fiducial CMB parameters.}
\label{alphabeta}
\end{center}
\end{figure}

Finally, we consider the case where the primordial perturbations $\chi_0(\vx)$ are modulated in real space, so that
the primordial perturbation field is
\be
\chi(\vx) = \chi_0(\vx)[1  + \phi(\vx)].
\ee
The primordial Gaussian field $\chi_0$ is assumed to be statistically homogeneous. Similar modulations have been considered before~\cite{Dvorkin:2007jp}.
We consider the modulating field $\phi(\vx)$ to be fixed, so the aim is to reconstruct the large-scale $\phi$ field by looking at the induced modulation of smaller-scale perturbations.
To leading order in $\phi$ the primordial covariance is given by
\begin{multline}
 \la  \chi(\vk)\chi(\vk')\ra =  (2\pi)^3 \delta(\vk+\vk') P_\chi(k) \\
+\int \ud^3 \vx e^{-i(\vk+\vk')\cdot\vx} \phi(\vx)\left[P_\chi(k) + P_\chi(k') \right].
\end{multline}
Note that the modulated field (with fixed $\phi$) is no longer statistically homogeneous. Expanding the exponentials using
\be
e^{i\vk\cdot\vx} = 4\pi \sum_{lm} i^l j_l(kx) Y_{lm}(\hat{\vx})Y_{lm}^*(\hat{\vk}),
\ee
and using Eq.~\eqref{TlmTransfer} to relate the primordial perturbations to the observed temperature multipoles, the covariance is then
\colswitch{ 
\ba
&& C_{l_1m_1 l_2 m_2} = \delta_{l_1 l_2}\delta_{m_1 m_2}C_{l_1}  +
\int \ud^3 \vx \phi(\vx) [\alpha_{l_1}(x) \beta_{l_2}(x) + \alpha_{l_2}(x) \beta_{l_1}(x)] Y_{l_1 m_1}^*(\hat{\vx})  Y_{l_2 m_2}(\hat{\vx})
\ea
}{ 
\ba
&& C_{l_1m_1 l_2 m_2} = \delta_{l_1 l_2}\delta_{m_1 m_2}C_{l_1}
\nn \\ && \quad +
\int \ud^3 \vx \phi(\vx) \alpha_{l_1}(x) \beta_{l_2}(x) Y_{l_1 m_1}^*(\hat{\vx})  Y_{l_2 m_2}(\hat{\vx})
\nn \\ && \quad +
\int \ud^3 \vx \phi(\vx) \alpha_{l_2}(x) \beta_{l_1}(x) Y_{l_1 m_1}^*(\hat{\vx})  Y_{l_2 m_2}(\hat{\vx}),\quad
\ea
}
where
\begin{eqnarray}
\alpha_{l}(r) &\equiv& 4\pi \int \ud \ln k j_l(kr) \frac{k^3\Delta_{l}(k)}{2\pi^2}\nn \\
\beta_{l}(r) &\equiv& 4\pi \int \ud \ln k j_l(kr) \Delta_{l}(k) \clp_\chi(k).
\label{alpha_beta_def}
\end{eqnarray}
Hence using Eq~\eqref{eqn:hest} there is a QML estimator for the modulating field $\phi$ with
\begin{multline}
\ph^{\phi}_{lm}(r) =
\int d\Omega Y_{lm}^{*}
\left[ \sum_{l_1 m_1} \alpha_{l_1}(r)\tf_{l_1 m_1} Y_{l_1 m_1} \right]
\\ \times
\left[ \sum_{l_2 m_2} \beta_{l_2}(r) \tf_{l_2 m_2} Y_{l_2 m_2} \right].
\label{localaniso}
\end{multline}
This allows us to reconstruct a map of the modulation at any desired radius. The construction is very similar to that used when estimating a local bispectrum~\cite{Komatsu:2003iq,Smith:2006ud}, where the main quantity of interest is the correlation of the reconstructed $\phi$ with an estimator of $\chi$ (see Appendix~\ref{bispectrum} for further discussion). By looking at the modulating field directly, we can also consider the case where $\phi$ and $\chi$ are uncorrelated,
and hence there is no leading-order bispectrum.

The isotropic component of the normalization is given by
\begin{multline}
\left[ \f^{\phi\phi}_{\rm iso}  \right]_{l m, l' m'}=
\delta_{ll'} \delta_{mm'} \times
\\
\sum_{l_1,l_2} \frac{(2l_1+1)(2l_2+1)}{8\pi}
\threejz{l}{l_1}{l_2} ^2 \frac{(\alpha_{l_1} \beta_{l_2} + \beta_{l_1} \alpha_{l_2} )^2}{\Ctot_{l_1} \Ctot_{l_2}},
\end{multline}
and typical $\alpha$ and $\beta$ functions are shown for adiabatic
modes in Fig.~\ref{alphabeta}.

Integrating $\ph^{\phi}_{lm}(r)$ over a window function $W(r)$
 is equivalent to estimating a modulating field with that separable radial dependence; for large scale modulations
 with a constant $W(r)$ this is close to the spatial modulation estimator considered in
 Sec.~\ref{sec:modulation_on_the_sky} since any large-scale primordial modulation modulates essentially
 all of the CMB anisotropy sources in that direction (except possibly low redshift sources due to ISW that only effect low $l$).
Mathematically this is because
\be
\int r^2 \ud r \alpha_{l_1}(r)\beta_{l_2}(r)\sim C_{l_1}
\ee
for $l_1$ close to $l_2$ (with equality for $l_1=l_2$).
We therefore do not repeat the analysis here, since the large-scale results of Sec.~\ref{sec:modulation_on_the_sky}
with the full range of $l$ can be interpreted as being approximately a constraint on large-scale primordial modulations.
We showed that any total modulation must be $\alt 1\%$ to be consistent the data on all scales, so modulation of primordial adiabatic
modes would not explain the marginal evidence for a $\clo(10\%)$ modulation at $l \alt 100$.

 We can also consider the case where there is a modulated primordial isocurvature mode, combined with the usual unmodulated adiabatic mode. For example the modulation could be due to a large-scale perturbation in the background curvaton field~\cite{Erickcek:2009at}. The off-diagonal covariance in general depends on  the power spectrum of the isocurvature modes as well as the correlation with the adiabatic modes, and the
isotropic power spectrum is also modified due to the extra contributions (which gives a constraint on the amplitude of the isocurvature contributions). Estimators for the modulating field are of the same form as Eq.~\eqref{alpha_beta_def} but with the appropriate combinations of isocurvature transfer functions and power spectra: either $\alpha$ and $\beta$ both calculated with isocurvature transfer function, or the cross-estimator
\begin{multline}
\ph^{\phi_{ai}}_{lm}(r) = \frac{1}{2}
\int d\Omega Y_{lm}^{*} \times \\
\biggl(
\left[ \sum_{l_1 m_1} \alpha^\iso_{l_1}(r)\tf_{l_1 m_1} Y_{l_1 m_1} \right]
 \left[ \sum_{l_2 m_2} \beta_{l_2}^\ad(r) \tf_{l_2 m_2} Y_{l_2 m_2} \right]
\\
+\left[ \sum_{l_1 m_1} \alpha^\ad_{l_1}(r)\tf_{l_1 m_1} Y_{l_1 m_1} \right]
\left[ \sum_{l_2 m_2} \beta_{l_2}^\iso(r) \tf_{l_2 m_2} Y_{l_2 m_2} \right]
\biggr).
\end{multline}
This corresponds to reconstructing the modulation of a primordial perturbation $\chi_0 + \phi S$, where $S$ is taken to be a CDM isocurvature perturbation correlated to $\chi_0$.
 We assume the primordial adiabatic, isocurvature and cross power spectra are proportional, so up to normalization $\alpha$ and $\beta$ differ only in the transfer function used.

\begin{figure}
\begin{center}
\epsfig{figure=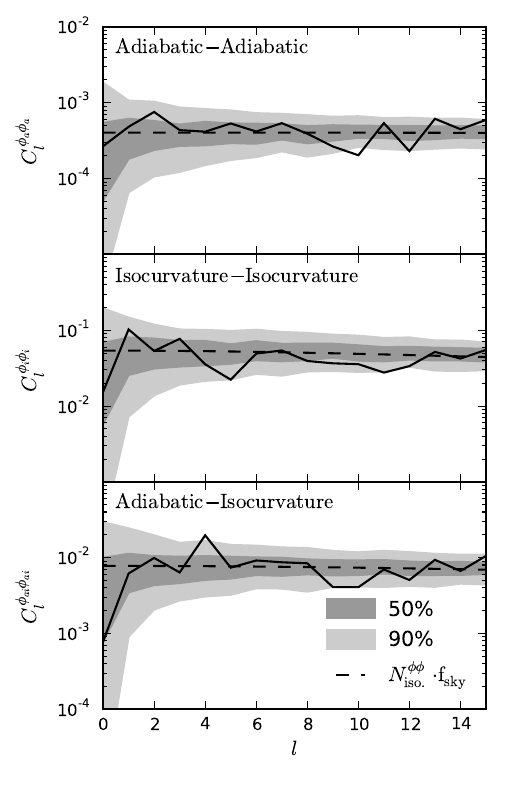,width=\columnwidth}
\caption{Pseudo-$C_{l}$ of the $\phi_{l m}$ reconstructions for the
  WMAP V-band foreground-reduced data, with KQ85 mask and $l_{\rm
    max}=400$. The full-sky, isotropic normalization appropriate to each estimator was used rather
  than the actual inverse Fisher matrix. The $[25,75]\%$ (dark gray)
  and $[5,95]\%$ (light gray) confidence intervals measured from
  Monte-Carlo simulations are overlaid. The $\phi$ estimates are
  scaled by $\Delta_r [r^{*}]^2$ to give a dimensionless constraint,
  with $\Delta_r = 100$Mpc, the approximate width of the last
  scattering surface.}
\label{fig:psmod_pseudo_cl}
\end{center}
\end{figure}

In Fig~\ref{fig:psmod_pseudo_cl} we plot the pseudo-$C_{\ell}$ power
spectrum of $\phi_{lm}$ reconstructed from the WMAP data for these
models. Large scale modulations should be approximately constant
across the depth of the last-scattering surface, and so we simply
evaluate
the estimator at the radius of peak visibility where most of the small-scale CMB fluctuations are coming from.
The estimators at a given $r$ are significantly more noisy than the sky modulation estimators since the latter
effectively average the result from many radial shells. We see no deviations from isotropy at the peak of the visibility.


As explained above, doing a constant line-of-sight integral is almost equivalent to a sky modulation for
large-scale modulations. The same applies for isocurvature mode
contributions (with greater accuracy since there is less contribution from ISW at low redshift), for example where the sky
has adiabatic and isocurvature contributions:
\be
\Theta(\vnhat) = \Theta^\ad(\vnhat) + \alpha[1+f(\vnhat)]\Theta^\iso(\vnhat).
\ee
The isocurvature power spectrum $C^{\iso-\iso}_l$ falls off on sub-horizon scales, so the data is consistent with an
 $\clo(10\%)$ dipole modulation of the total temperature if the modulation is only in the isocurvature component.
 The isotropic power spectrum constrains the isocurvature component to be subdominant however, so this would mean that an $\clo(1)$ modulation
 of the isocurvature component is required, which would invalidate the assumptions in deriving the sky modulation estimator that $|f|\ll 1$.
 However we can still apply a null-hypothesis test on a purely isotropic adiabatic model using whatever parameterization we like,
 though interpretation of any detection of deviation may need to be treated with care. We do not attempt a fully consistent analysis
 of isotropic and anisotropic constraints on isocurvature contributions here.

The isocurvature mode could also be correlated with the adiabatic mode,
in which case we can constrain a model of the form
\be
\Theta(\vnhat) = \Theta^\ad(\vnhat) + f(\vnhat)\Theta^\iso(\vnhat),
\ee
where $\la\Theta_{lm}^\ad\Theta^{\iso *}_{lm}{}\ra = \alpha C^{\ad-\iso}_l$, $\alpha$ depends on the amplitude and degree of correlation,
 and $C^{\ad-\iso}_l$ is the
spectrum for totally correlated modes with the same initial amplitude ($S=-\chi_0$). As usual we neglect terms of $\clo(f^2)$. The monopole part of $f$ allows for an isotropic
isocurvature contribution.
The estimators are almost the same as in the total temperature modulation case, with
\be
\ph^{f}_{lm} =
\int d\Omega Y_{lm}^{*} \left[ \sum_{l_1 m_1}^{l_{\rm max}} \tf_{l_1 m_1} Y_{l_1 m_1} \right] \left[ \sum_{l_2 m_2}^{l_{\rm max}} C^X_{l_2} \tf_{l_2 m_2} Y_{l_2 m_2} \right],
\ee
where $C^X_l$ is the relevant power spectrum, and we evaluate $\tf_{lm}$ using the standard best-fit theory model in the assumption that
a model with isocurvature modes must have about the same spectrum as the null hypothesis with no isocurvature modes.
A more general model could also allow part of the adiabatic mode to be modulated.

\begin{figure}
\begin{center}
\epsfig{figure=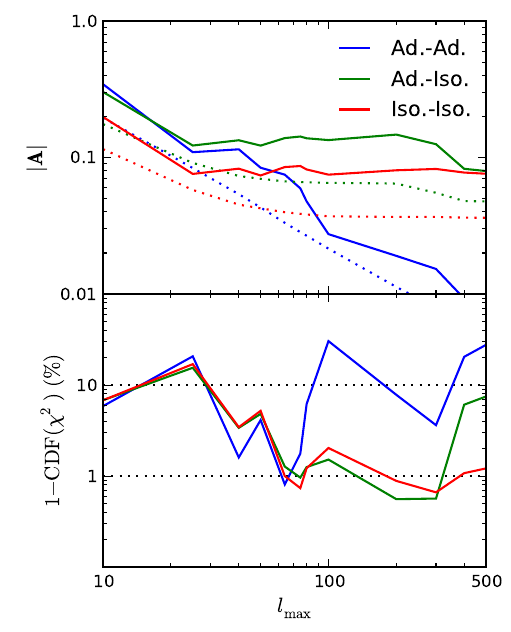,width=\columnwidth}
\caption{Same as Fig.~\ref{fig:modulation_ml_QVW_std_AS} but using the
  power spectra for the indicated combinations of adiabatic and
  isocurvature transfer functions in the estimator (replacing
  $C_{l_2}$ in Eq.~\eqref{modulation_h}). All curves are for the WMAP V-band foreground-reduced data with KQ85 masking.}
\label{fig:iso_modulation}
\end{center}
\end{figure}

 The constraint on isocurvature modulations is shown in Fig.~\ref{fig:iso_modulation},
though the data is still consistent with isotropy at the $1\%$-level. See Ref.~\cite{Erickcek:2009at} for
more detailed discussion and possible physical mechanisms for generating modulated isocurvature modes.

We have focussed on local modulations here, however the formalism can in principle be applied to more general primordial anisotropic models, for instance with
\begin{multline}
\chi(\vx) = \chi_0(\vx) +\\ \int \frac{\ud^3 \vk_1}{(2\pi)^3} \frac{\ud^3 \vk_2}{(2\pi)^3} f(\vk_1, \vk_2) \chi_0(\vk_1)\phi(\vk_2) e^{i(\vk_1+\vk_2)\cdot \vx}.
\end{multline}
Such more general forms can arise for example in warm inflation~\cite{Moss:2007cv} and other models giving non-local primordial bispectrum non-Gaussianity (see e.g. Ref.~\cite{Fergusson:2008ra}). A detailed analysis is beyond the scope of this paper, but is similar to that required to constrain the bispectrum when $\phi$ is equal to $\chi_0$.

\section{Conclusions}
The assumption that the CMB is statistically isotropic
is a common one in modern cosmology, and needs
to be rigorously tested.
In this paper, we argued that QML estimators
are often ideal for this purpose. They provide optimal, minimum-variance
constraints and
permit straightforward systematic and jacknife tests.
We have demonstrated their use in application to the WMAP dataset
for constraining several anisotropic models.
Most notably, we reproduced the results of previous exact likelihood analyses
which showed marginal evidence for a dipolar modulation of the
observed CMB for $l < 100$. However we showed that the amplitude of the modulation must fall significantly on smaller scales,
consistent with observations that the small scale
power in the universe (e.g. probed by quasars~\cite{Hirata:2009ar})
appears to be highly isotropic. This scale-dependence is inconsistent with modulated adiabatic primordial modes,
but could possibly be explained by modulated isocurvature modes, which have less power on small scales. However
evidence for deviations from isotropy are only at the $1\%$-level.

We also studied a model in which the primordial power spectrum
has a direction dependence, detecting a significant anisotropy in
the WMAP maps with close-to-ecliptic alignment.
However we argued that this signal has an important contribution from
beam asymmetries, which are uncorrected in the maps, and that
significant variations between differencing assemblies indicate that most of the signal is non-primordial.

An important extension of this approach is the inclusion of
polarization data. For anisotropies which are detected in temperature
at marginal significance, polarization can help to increase the
signal  to noise
and to distinguish possible primordial, local and systematic origins
of any detection~\cite{Dvorkin:2007jp}.

\section{Acknowledgements}
 AL acknowledges a PPARC/STFC advanced fellowship and thanks the
Aspen Center for Physics for hospitality while part of this work in
progress, and workshop participants for useful discussions.
DH is grateful for the support of a Gates scholarship.
We thank Anthony Challinor for helpful suggestions and Hiranya Peiris, Andrew Pontzen, Kendrick Smith and
Douglas Scott for numerous related discussions.
Some of the results in this paper have been derived using
\healpix~\cite{Gorski:2004by}.
We acknowledge the use of the Legacy Archive for Microwave Background
Data Analysis (LAMBDA).
Support for LAMBDA is provided by the NASA Office of Space Science.

\appendix

\section{Anisotropic Simulations}
\label{sec:anisotropic_simulations}
Here we present an algorithm for generating weakly anisotropic simulations\footnote
{Thanks to Anthony Challinor for suggesting this presentation.}.
We wish to sample from a Gaussian distribution with covariance
\be
\mC = (\mI + \delta\mC[\mC^{i}]^{-1})\mC^i
\ee
where $\mC^i$ is isotropic and $\delta\mC$ is the small anisotropic part. Then
\be
\vTheta = \left[\mI + \delta\mC[\mC^{i}]^{-1}\right]^{1/2}\vTheta^{i},
\ee
where $\vTheta^i$ has covariance $\mC^{i}$ will have the required covariance.
For small anisotropy this can be expanded to give
\be
\vTheta \approx \vTheta^{i} + \frac{1}{2}\delta\mC[\mC^{i}]^{-1}\vTheta^{i} + \dots.
\ee
Higher order terms can be included if required; here we just calculate the leading anisotropic term.
In the case of an anisotropic primordial power spectrum this is
\be
\delta\vTheta_{lm} \approx
\frac{1}{2}
\int \ud\Omega Y_{l m}^*
g \left[\sum_{l_2 m_2} i^{l-l_2}\frac{C_{l l_2}\Theta^{i}_{l_2 m_2}}{C_{l_2}}Y_{l_2 m_2} \right].
\ee

\section{Relation to bispectrum estimators}
\label{bispectrum}
Power anisotropy is closely related to primordial non-Gaussianity.  Local non-Gaussianity gives power in squeezed triangles: large scale perturbations are correlated to small scale power over their extent.
 If the large scale perturbation were unknown, this would look just like a small-scale power anisotropy. Hence a method that reconstructs the power
 anisotropy should be sensitive to three-point non-Gaussianity, in that the correlation of the anisotropy estimator with the temperature is a probe of the bispectrum. Non-Gaussianity can also give rise to a dipole power asymmetry correlated with the (unknown) cosmological dipole. In the inflationary context a power anisotropy on a given scale typically indicates the impact of those modes when they were outside the horizon on the generation of smaller fluctuations, e.g. via the local change in the effective background seen by the smaller modes as they leave the horizon.
  Unfortunately the large variance of the anisotropy estimators means that for small amounts of non-Gaussianity any detection with standard shapes is expected to be via the correlation to the temperature, as in standard bispectrum analyses. However since there is marginal evidence for power anisotropies in the data it is worth considering whether these could be associated with some form of physically-motivated primordial non-Gaussianity. There may also be marginal observational evidence for local primordial bispectrum non-Gaussianity~\cite{Yadav:2007yy,Komatsu:2008hk,Smith:2009jr}, so conceivably there could be a joint explanation.

A statistically-isotropic and parity-invariant CMB bispectrum $B_{l_1 l_2 l_3}$ is defined by
\begin{eqnarray}
\la a_{l_1 m_1} a_{l_2 m_2} a_{l_3 m_3}\ra &\equiv&
B_{l_1 l_2 l_3}\threej{l_1}{l_2}{l_3}{m_1}{m_2}{m_3}\\
&=& b_{l_1 l_2 l_3} \int \ud\Omega Y_{l_1 m_1} Y_{l_2 m_2} Y_{l_3 m_3},\qquad
\end{eqnarray}
where $b_{l_1 l_2 l_3}$ is the reduced bispectrum. When the CMB power spectrum is known, the optimal estimator for weakly non-Gaussian fields with bispectrum $B_{l_1 l_2 l_3}$ is~\cite{Creminelli:2005hu,Babich:2005en,Smith:2006ud}
\begin{multline}
\cle = N_\cle^{-1}\sum_{l_i m_i}
 B_{l_1 l_2 l_3} \threej{l_1}{l_2}{l_3}{m_1}{m_2}{m_3} \\\times \biggl[
\bTheta_{l_1 m_1} \bTheta_{l_2 m_2}\bTheta_{l_3 m_3}
- 3 C^{-1}_{l_1 m_1 l_2 m_2} \bTheta_{l_3 m_3}\biggr],
\end{multline}
where $N_\cle$ is a normalization. Following Ref.~\cite{Smith:2006ud} this can be written as
\be
\cle = \frac{1}{N_\cle} \sum_{l m} \bTheta_{l m}^*( X_{lm} - 3\la X_{lm}\ra)
\ee
where
\begin{eqnarray}
X_{lm} &=& \sum_{l_1 m_1, l_2 m_2} B_{l l_1 l_2} (-1)^{m_1}\threej{l}{l_1}{l_2}{m}{-m_1}{m_2} \bTheta_{l_1 m_1} \bTheta_{l_2 m_2}^* \nn\\
&=&\int \ud\Omega Y_{l m}^{*} \times \nn\\
&&\sum_{l_1 l_2} b_{l l_1 l_2}
\left[ \sum_{m_1} \tf_{l_1m_1} Y_{l_1m_1} \right] \left[ \sum_{m_2} \tf_{l_2m_2} Y_{l_2m_2} \right].
\end{eqnarray}
The $X_{lm}$ field is of the form of the most general quadratic anisotropy estimator, with free weighting coefficients $b_{l l_1 l_2}$. Bispectrum estimators are essentially correlations of specific power anisotropy estimators with the temperature.

Often the dipole $\Theta_{1m}$ is projected out of bispectrum analyses, effectively ignoring $X_{1m}$, even though in general $X_{lm}$ contains interesting power asymmetry information that could be generated by non-Gaussianity. The `optimal' bispectrum estimator is only optimal with certain assumptions; if the temperature dipole information is removed then the dipole part of the power
anisotropy should contain additional information about the non-Gaussianity, which in principle could improve constraints if the non-Gaussianity is larger
than the anisotropy expected from lensing and doppler effects. This is equivalent to using part of the trispectrum, see Refs.~\cite{Kogo:2006kh,Creminelli:2006gc} for further discussion.


A primordial statistically isotropic bispectrum can be defined as
\be
\la  \chi_0(\vk_1)\chi_0(\vk_2)\chi_0(\vk_3)\ra =  (2\pi)^3 \delta(\vk_1+\vk_2+\vk_3) B_\chi(k_1,k_2,k_3).
\ee
For local non-Gaussianity
\be
B_\chi(k_1,k_2,k_3) = \pm 2 \frac{3}{5}f_{NL}(P_\chi(k_1) P_\chi(k_2) + \text{2 perms.})
\ee
where the $3/5$ is conventional (from relation the curvature perturbation to the Newtonian potential in the
radiation dominated era) and $\pm$ is a sign convention. The reduced bispectrum is then
\be
b_{l_1 l_2 l_3} = \pm\frac{3}{5}f_{NL}\int r^2\ud r  \beta_{l_1}(r) \beta_{l_2}(r) \alpha_{l_3}(r) + \text{5 perms.}
\ee

We can relate this to our local spatial modulation estimator of Eq.~\ref{localaniso} by windowing to give a line-of-sight average of the primordial modulating field
\begin{multline}
\bar{h}_{lm} = \int \ud r r^2 \ph_{lm}^\phi(r) W_l(r)
=\int \ud\Omega Y_{l m}^{*} \\
\times\sum_{l_1 l_2} b_{l l_1 l_2}\left[ \sum_{m_1} \tf_{l_1m_1} Y_{l_1m_1} \right] \left[ \sum_{m_2} \tf_{l_2m_2} Y_{l_2m_2} \right].
\end{multline}
This is of the form of a general quadratic anisotropy estimator, but with weights in the specific form
\be
b_{l l_1 l_2} = \int \ud r r^2 W_l(r) \alpha_{l_1}(r)\beta_{l_2}(r).
\ee
If the modulating field is the primordial field itself, $\phi = \chi_0$, there is local bispectrum. In this case choosing
$W_l(r) = \beta_l(r)$ relates the primordial field to the observed temperature, since $\beta(r)\tf_{lm}$
is the minimum variance estimate for the primordial field at $r$~\cite{Komatsu:2003iq}. Then
$b_{l l_1 l_2}$ is the form of the reduced bispectrum obtained by correlating $\bar{h}_{lm}$ with the temperature $\tf_{lm}$.

\bibliography{aniso}

\end{document}